# Theory on the mechanism of DNA renaturation: Stochastic nucleation and zipping

Gnanapragasam Niranjani and Rajamanickam Murugan*
Department of Biotechnology, Indian Institute of Technology Madras
Chennai 600036 India

* rmurugan@gmail.com

**ABSTRACT**
Renaturation of complementary single strands of DNA is one of the important processes that requires better understanding in the view of molecular biology and biological physics. Here we develop a stochastic dynamical model on the DNA renaturation. According to our model there are at least three steps in the renaturation process viz. incorrect-contact formation, correct-contact formation and nucleation, and zipping. Most of the earlier two-state models combined nucleation with incorrect-contact formation step. In our model we suggest that it is considerably meaningful when we combine the nucleation with the zipping since nucleation is the initial step of zipping and the nucleated and zipping molecules are indistinguishable. Incorrect-contact formation step is a pure three-dimensional diffusion controlled collision process. Whereas nucleation involves several rounds of one-dimensional slithering dynamics of one single strand of DNA on the other complementary strand in the process of searching for the correct-contact and then initiate nucleation. Upon nucleation, the stochastic zipping follows to generate a fully renatured double stranded DNA. It seems that the square-root dependency of the overall renaturation rate constant on the length of reacting single strands originates mainly from the geometric constraints in the diffusion controlled incorrect-contact formation step. Further the inverse scaling of the renaturation rate on the viscosity of the reaction medium also originates from the incorrect-contact formation step. On the other hand the inverse scaling of the renaturation rate with the sequence complexity originates from the stochastic zipping which involves several rounds of crossing over the free-energy barrier at microscopic levels. When the sequence of renaturing single strands of DNA is repetitive with less complexity then the cooperative effects will not be noticeable since the parallel zipping will be a dominating factor. However for DNA strands with high sequence complexity and length one needs to consider the cooperative effects both at microscopic and macroscopic levels to explain various scaling behaviours of the overall renaturation rate.

# INTRODUCTION

The biological function of DNA depends largely on its double stranded helical structure and its ability to unwind and rewind in a reversible manner. The double stranded structure of DNA (dsDNA) is mainly stabilized by weak hydrogen bonds between the nitrogen bases of complementary single strands (c-ssDNAs) and the hydrophobic forces arising from the base-stacking within the core of dsDNA polymer (**1-2**). These weak interactions melt down upon heating the solution containing dsDNA beyond the melting temperature which in turn yields the corresponding c-ssDNAs. These single strands exactly reunite (hybridize) back into their original double stranded helical form upon cooling the solution below the melting temperature. Melting temperature of dsDNA is defined as the temperature at which precisely half of the dsDNA melts into corresponding c-ssDNAs. Several molecular biological processes such as transcription, translation and replication and *in vitro* laboratory techniques are solely based on the reversible winding-rewinding property of dsDNA. Understanding the dynamics and mechanism of renaturation of c-ssDNAs in solution is important in recombination, design of primers for polymerase chain reaction, design of oligonucleotide probes for microarray chips, various membrane blotting techniques and other related DNA fingerprinting technologies (**2-5**). In this context, several models describing the process of renaturation of c-ssDNAs in aqueous solution have been developed and experimentally verified (**6–29**). Detailed understanding of the mechanism of renaturation of c-ssDNA at microscopic level is one of the important contemporary topics of interest in molecular biology and biological physics.

Renaturation of c-ssDNAs was initially thought (**6**) as one-step bimolecular second order chemical kinetic process. Several experimental observations could not be explained by a simple one-step second order kinetics. One of such observations is that irrespective of the experimental conditions the overall bimolecular rate constant was directly proportional to the square-root of the average length of c-ssDNAs and inversely proportional to its sequence complexity. Whereas a one-step process predicted that the bimolecular collision rate was directly proportional to the average length of c-ssDNA. To comply with various experimental observations, Wetmur and Davidson (**6**) suggested a detailed two-step renaturation model that comprised of a nucleation and zipping steps. They had shown that the overall second order rate constant associated with the renaturation phenomenon could be expressed as function of average length of the sheared DNA and complexity of the reacting c-ssDNAs. Here complexity is defined as the length of DNA stretch with a unique nucleotide sequence pattern. Wetmur and Davidson further formulated a theoretical model on renaturation phenomenon according to which the overall bimolecular rate constant was directly proportional to the nucleation rate apart from the ratio of average length of c-ssDNA and complexity. They argued that the nucleation rate constant is inversely proportional to the square-root of average length of c-ssDNA so that the overall bimolecular rate is directly proportional to the square-root of length of c-ssDNA as observed in experiments. They suggested that the inverse scaling of nucleation rate constant with length must be owing to either the thermodynamic excluded volume effects associated with the intra-strand dynamics or stearic hindrance associated with the diffusion controlled interpenetration of c-ssDNAs that is essential for the nucleation step.

Here one should note that the two-step model of Wetmur-Davidson will be inconsistent whenever the complexity has same magnitude as that of the length of c-ssDNA. Subsequent experimental studies on renaturation phenomenon were mainly focussed (**16-28**) on unravelling the molecular mechanisms and the underlying thermodynamics and kinetics aspects. In line of these experimental studies, several theoretical and computational models (**6**, **7**, **14-17**, **23**-**27**) were also developed to explain the observed scaling behaviours of the overall second order renaturation rate constant on the size of reacting c-ssDNAs, temperature, ionic strength and viscosity of the reaction medium. Recent studies (**23**-**27**) considered either the excluded volume effects acting on intrastrand dynamics or stearic hindrance associated with the diffusion controlled interpenetration of c-ssDNAs to explain the observed scaling behaviours of the overall renaturation rate. Diffusion based models provide correct viscosity dependence of overall renaturation rate constant compared to the models based on the framework of transition state theories (TST). Recently nucleation step in renaturation was modelled as an escape over free energy barrier (**27**) within the framework of Kramer's theory that deals with the dynamics of Brownian particle over a potential energy barrier. It was argued (**27**) that the square root dependency of nucleation rate on the length of c-ssDNA mainly originates from the entropic component of the free energy barrier associated with the Kramer's escape problem. Though this approach appeared to be reasonable, nature of the reaction coordinate and potential energy barrier associated with the renaturation process were not clearly defined. Moreover the exact connection between the entropic component of the free energy barrier and the observed scaling behaviour was not clearly established in detail.

According to the current theoretical understandings over experimental and computational observations (**23**-**29**), the renaturation process should have at least three distinct steps namely (a) incorrect contact formation (b) nucleation or correct contact formation and (c) zipping. In the first step, the reacting c-ssDNAs collide with each other via three-dimensional (3D) diffusion controlled routes. This results in the formation of Watson-Crick (WC) base pairs at random incorrect contacts between the reacting c-ssDNAs. Such incorrect WC contacts translocate along c-ssDNAs either via one-dimensional (1D) slithering dynamics or internal displacement (**29**) mechanisms until finding the correct-contact and initiate the nucleation process which is in turn followed by spontaneous zippering of c-ssDNAs. Here one should note that steps (a) and (b) are purely stochastic dynamical processes similar to that of the site-specific DNA-protein interactions (**17**, **30-31**). Moreover zippering step (c) too will be a stochastic process under weak renaturation conditions. These mean that one needs to apply stochastic dynamics based arguments rather than merely thermodynamics based ones to explain the observed scaling behaviours and underlying mechanisms. In this paper we will formulate such a stochastic dynamics based theoretical framework of renaturation phenomenon and explain various scaling properties associated with the renaturation rate.

## RESULTS
### Theoretical formulation of DNA renaturation kinetics
The basic steps of renaturation of c-ssDNA viz. (a) incorrect contact formation (b) nucleation or correct contact formation and (c) zippering can be represented by a schematic reaction **Scheme I** in **Fig.1**. Here [$S$] and [$S'$] are the concentrations (mol/lit, M) of the colliding c-ssDNAs whose lengths are $L$ bps (template) and $l$ bps (probe) respectively where by definition $L \geq l$, [$X$] is the concentration of c-ssDNAs with incorrect contacts (ci-ssDNA), [$Y_N$] is the concentration of c-ssDNAs with correct-contact (cc-ssDNA) and nucleation, and [$Z$] is the concentration of completely renatured dsDNA form. Here one should note that [$Y_N$] is a special form of [$Z$] since the complementary strands are already nucleated and aligned in [$Y_N$]. Since it is very difficult to identify and quantity [$Y_N$] in experiments we consider [$Z$] as the main product of renaturation in this paper. Further $k_{fQ}$ (M$^{-1}$s$^{-1}$) and $k_r$ (s$^{-1}$) are the forward second-order and reverse first-order rate constants associated with the formation of incorrect contacts between colliding c-ssDNAs, $k_N$ (s$^{-1}$) is the nucleation rate constant and $k_Z$ (s$^{-1}$) is the zippering rate constant. Since [$Y_N$] is a hidden intermediate we consider the overall rate constant associated with both nucleation and subsequent spontaneous zipping as $k_{NZ}$ which is the inverse of total time required for nucleation and zipping processes together i.e. $k_{NZ} = 1/(1/k_Z + 1/k_N)$. Typical values of the size of nuclei (**6**) seems to be $N \sim 4$-$7$ bases. Here we set the subscript $Q = C$ for condensed conformational state of c-ssDNA and $Q = G$ for relaxed conformational state of c-ssDNA. The set of differential rate equations associated with **Scheme I** can be written as follows.

$$d[X]/dt = k_{fQ}[S][S'] - (k_r + k_{NZ})[X]; \ d[Z]/dt = k_{NZ}[X]; \ d[Z]/dt \simeq \bar{k}_S [S][S'] \quad (1)$$

Here we have defined the overall observed second order rate constant $\bar{k}_S = k_{fQ} k_{NZ} / (k_{NZ} + k_r)$. In deriving the expression for $\bar{k}_S$ we have assumed that the nucleation and zipping are the rate limiting ones so that $d[X]/dt \simeq 0$ in the timescales of nucleation and zipping. Essentially the first reaction in **Scheme I** can be thought as collisions between spatially distributed clusters of nitrogen bases corresponding to two reacting c-ssDNAs as in case of a mean field approach (**27**). As a result, the overall bimolecular rate associated with the formation of incorrect contacts between spatially distributed base-clusters of c-ssDNAs is proportional of the product of concentrations of the total nitrogen bases in c-ssDNA molecules. The cylindrical surface area $S_Q \sim 2\pi r_D Q$ of a c-ssDNA molecule will be confined within the spherical solvent shell with surface area $SS_Q \simeq 4\pi r_Q^2$ ($Q = L$ for template and $Q = l$ for probe) where $r_Q$ is the radius of gyration of the respective c-ssDNA molecule. Under strongly condensed conformational state of c-ssDNA one finds that $SS_Q < S_Q$ and when the DNA polymer is in a relaxed conformational state then one find that $SS_Q > S_Q$. At a coarse grained level one can model the bases of c-ssDNA as a chain of spherical beads with radius $r_D$. Under relaxed conformational state all these nitrogen base beads are distributed on the surface of the spherical solvent shell that covers a c-ssDNA molecule (**Fig. 1B**). Whereas under condensed conformational state of c-ssDNA molecules significant fraction of nitrogen base beads will be inaccessible to the inflowing c-ssDNA molecules since they are buried inside the matrix of condensed c-ssDNA (**Fig. 1C**).

## Calculation of incorrect-contact formation rate

Actually most of the theoretical models derived the scaling over the length of c-ssDNAs mainly from the fact that the radius of gyration (measured in bps) associated with reacting c-ssDNAs scales with their length as $r_Q \propto Q^\alpha$ ($Q = L$ or $l$). Here the value of the exponent $0 < \alpha < 1$ varies depending on the type of polymer and solvent conditions. For an ideal Gaussian chain polymer that is immersed in a theta solvent (**32-33**) one finds that $\alpha \sim \frac{1}{2}$. Noting these facts the Smolochowski type limiting rate for a diffusion controlled incorrect contact forming step can be given as follows.

$$k_{fQ} = 4\pi r_Q^2 J_Q; \ J_Q = (D_l + D_L)/r_Q; \ D_S = k_B T / 6\pi \eta r_S; \ Q = \{C, G\}; \ S = \{L, l\} \quad (2)$$

Here $J_Q$ is the inflowing flux of c-ssDNA molecules, $D_S$ is the three-dimensional (3D) diffusion coefficient associated with the colliding c-ssDNAs and $r_Q$ is the reaction radius that depends on the conformational state of c-ssDNA. When c-ssDNA is relaxed, then all the base beads will be distributed over the surface of the solvent shell that covers the entire c-ssDNA polymer. Since $SS_Q > S_Q$ there will be several patches on the solvent shell surface without base beads. Noting that an incorrect contact can be formed only upon collision between probe and base beads of template, one needs to integrate over the surface of the template c-ssDNA polymer that is spread on the surface of spherical solvent shell rather than the entire surface of spherical solvent shell (**Fig. 1B**). On the other hand, when the c-ssDNA polymer is highly condensed then significant fraction of base beads will be buried inside the matrix of condensate (**Fig. 1C**). Therefore depending on the conformational state of colliding c-ssDNA molecules the bimolecular collision rate associated with the formation of incorrect contact between template and probe molecules can be written as follows.

Case I: Relaxed conformational state of c-ssDNA

$$k_{fG} \simeq 2\pi r_G L J_G; \; J_G \simeq (D_l + D_L)/r_G; \; r_G = r_D + r_l; \; \therefore k_{fG} \simeq k_t L(1/r_l + 1/r_L)/8 \tag{3}$$

Case II: Condensed conformational state of c-ssDNA

$$k_{fC} \simeq 4\pi r_C^2 J_C; \; J_C \simeq (D_l + D_L)/r_R; \; r_C = r_L + r_l; \; \therefore k_{fC} \simeq k_t (r_L + r_l)^2 / 4 r_L r_l \tag{4}$$

Here $k_t \simeq (8 k_B T / 3\eta)$ is the Smolochowski type diffusion controlled bimolecular collision rate limit (M$^{-1}$s$^{-1}$) where $k_B$ is the Boltzmann constant, $\eta$ is the viscosity of the reaction medium and $T$ is the temperature in degrees $K$ and the colliding molecules are same in size. In general one can write $k_{fQ} \simeq k_t \delta_Q$ where $\delta_G \simeq L(1/r_l + 1/r_L)/8$ and $\delta_C \simeq (r_L + r_l)^2 / 4 r_L r_l$. When the colliding c-ssDNA are same in size then one finds that $\delta_G \simeq L/4 r_L$ and $\delta_C \simeq 1$.

## Role of electrostatic repulsions at DNA-DNA interface

However in **Eqs. 3-4** we have not considered the electrostatic repulsions between the negatively charged phosphate backbones of c-ssDNA chains and shielding effects of solvent and other ion molecules at the DNA-DNA interface of c-ssDNA molecules. Upon considering this fact and following the detailed works of Montroll in **Ref. 34** we find the expression for the modified bimolecular rate constants as follows.

$$k_{fQ} = k_t \psi_Q \delta_Q; \; \psi_Q \simeq (\kappa / r_Q) / (e^{\kappa / r_Q} - 1); \; \kappa = z_S z_{S'} e^2 / \zeta k_B T; \; Q = (C, G) \tag{5}$$

Here $\kappa$ is the Onsager radius which is defined as the distance between negatively charged phosphate backbones of colliding c-ssDNA chains at which the electrostatic repulsive energy is same as that of the thermal energy (~$k_B T$), $z_S e$ and $z_{S'} e$ are the overall charges on the respective c-ssDNA molecules. Since $|\kappa| \geq r_Q$ by definition and $\kappa > 0$ in the present context, one finds that $\psi \sim |\kappa| / r_Q$ for large values of $|\kappa|$ and for $|\kappa| = r_Q$ one obtains $\psi \simeq 1.58$. One should also note that $\psi_Q \simeq 1$ only when $|\kappa| = 0$. Nevertheless while deriving **Eq. 5** we have not considered the shielding effects of solvent ions over the electrostatic repulsive forces between the phosphate backbones of c-ssDNA. Upon following the Debye theory of kinetic salt effects over diffusion controlled rate processes (**35**), one can rewrite the modified bimolecular rate constant in the presence of other ions in the solvent as follows.

$$k_{fQ} \simeq k_t \chi_Q \delta_Q; \; \chi = \psi_Q e^{\Phi}; \; \Phi = 2 A z_S z_{S'} \sqrt{E} \tag{6}$$

Here $A = 0.509$, $E$ is the overall ionic strength of the medium and $Q = (C, G)$ as defined in **Eqs. 3** and **4** depending on the type of conformational state of c-ssDNA polymer.

## Calculation of nucleation time

We learnt from recent computational studies (**29**) that the colliding c-ssDNA with incorrect contacts between them undergo several trials of slithering and internal displacement dynamics before reaching the correct-contact and then nucleate the zipping process. These dynamical processes are similar to that of the facilitating 1D diffusional dynamics as in case of site specific DNA-protein interactions. Unlike the overall electrostatic attractive forces acting at the DNA-protein interface, in case of DNA renaturation there is a strong electrostatic repulsive force acting at the interface of ci-ssDNA that will be shielded by the solvent molecules present at the interface of ci-ssDNA strands. With this background one can model the slithering dynamics as 1D diffusion of one c-ssDNA molecule on the other in the process of searching for the correct-contact. To find the correct-contact on one c-ssDNA the other c-ssDNA needs to try out at least $\lambda = L/n$ stretches of 1D slithering with an average size of $n$ bases. This will ensure that the initial incorrect contact visits all the possible positions and subsequently the correct contact is formed. The mean first passage time ($\tau_N$) associated with the visit of all the possible positions of c-ssDNAs by the initial incorrect-contact between them via 1D diffusion dynamics can be given as $\tau_N = \lambda \bar{\tau}_c$ where $\bar{\tau}_c \simeq n^2/12D_o$ is the average time that is required by a unbiased 1D random walker to visit $n$ sites of a linear lattice. This result can be obtained as follows. The stochastic differential equation associated with an unbiased random walker on a linear lattice can be written as follows (**36-38**).

$$dx/dt = \sqrt{2D_o}\Gamma_t; \; x \in (0,n); \; \langle \Gamma_t \rangle = 0; \; \langle \Gamma_t \Gamma_{t'} \rangle = \delta(t-t') \tag{7}$$

Here $D_o$ is the one dimensional diffusion coefficient associated with the dynamics of the random walker, $\Gamma_t$ is the Gaussian white noise with mean and variance as given in **Eq. 7**. The probability density function associated with the dynamics of such random walker obeys the following forward Fokker-Planck equation (FPE) with initial and boundary conditions.

$$D_o\left(\partial^2 p_{x,t}/\partial x^2\right) = \partial p_{x,t}/\partial t; \; p_{0,t} = p_{n,t} = 0; \; p_{x_0,0} = \delta(x-x_0) \tag{8}$$

The mean first passage time (MFPT) associated with the escape of a random walker from the interval $x \in (0,n)$ starting from an arbitrary lattice point $x$ inside the interval obeys the following backward Fokker-Planck equation with boundary conditions as in **Eq. 8**.

$$D_o\left(d^2\tau_x/dx^2\right) = -1; \; \tau_0 = \tau_n = 0; \; \tau_x = \left(n^2 - x^2\right)/2D_o; \; \bar{\tau}_c = (1/n)\int_0^n \tau_x dx \simeq n^2/12D_o \tag{9}$$

Since the random walker can enter initially anywhere in interval $x \in (0,n)$ of linear lattice with equal probabilities one needs to average the computed MFPT $\tau_x$ over all values of initial positions $x$ and as in **Eq. 9** we find the initial position averaged value as $\bar{\tau}_c \simeq n^2/12D_o$. This is approximately the time that is required by a random walker to visit all the sites of a linear lattice confined inside the interval $x \in (0,n)$ starting from anywhere inside the interval.

## Calculation of zipping time

Formation of correct-contact will result in the nucleation of zipping process. Upon formation of a nucleation site, the subsequent stochastic zipping of cc-ssDNA can be described by the following birth-death master equation.

$$\partial_t P(u,t) = k_+ P(u-1,t) + k_- P(u+1,t) - (k_+ + k_-) P(u,t) \tag{10}$$

Here $P(u, t)$ is the probability of finding the cc-ssDNA with $u$ number of correct contacts at time $t$ starting from the nucleation, $k_+$ (s$^{-1}$) and $k_-$ (s$^{-1}$) are the respective average forward and reverse rate constants associated with the microscopic zipping reaction. Here the initial and boundary conditions corresponding to **Eq. 10** can be written as follows.

$$P(u,t_0) = P(u,t_0 | u_0, t_0) = \delta(u - u_0); \; k_- P(1,t) = k_+ P(0,t); \; P(\beta+1, t) = 0 \tag{11}$$

The mean first passage time associated with the complete zipping obeys the following backward type master equation with similar boundary conditions.

$$k_+ U(u) - k_- U(u-1) = -1; \; U(u) = \tau(u+1) - \tau(u); \; \tau(\beta+1) = 0; \; \tau(-1) = \tau(0) \tag{12}$$

Here $u = 1$ is a reflecting boundary and $u = \beta$ is the absorbing boundary. One can solve the difference equation **Eq. 10** as follows. By defining equilibrium constant as $K_R = (k_-/k_+)$, **Eq. 10** can be rewritten in the following form.

$$k_+ u \phi(u) [S(u) - S(u-1)] = -1; \; \phi(u) = \prod_{z=2}^{u} K_R; \; S(u) = U(u)/\phi(u)$$

Upon solving this difference equation for the boundary conditions given in **Eq. 11** we find the following expression for the overall mean first passage time associated with complete zipping of $\beta$ correct contacts ($u = \beta$) of cc-ssDNA starting from the number of correct contacts $u = 1$.

$$\tau_Z = \sum_{u=1}^{\beta} \phi(u) \sum_{w=1}^{u} (k_+ \phi(w))^{-1} = \left( K_R^{\beta+1} - K_R(\beta+1) + \beta \right) / k_+ (1 - K_R)^2 \tag{13}$$

From this equation we find the limits $\lim_{K_R \to 1} \tau_Z \simeq \beta(1+\beta)/2k_+$ and $\lim_{K_R \to 0} \tau_Z \simeq \beta/k_+$ where $\beta = L/l_p$ is a dimensionless quantity which is the total number of correct-contacts between colliding c-ssDNA upon formation of dsDNA. Here $L$ is total length of c-ssDNA in bases and $l_p = 1$ base (1 base $\sim 3.4 \times 10^{-10}$ m). When the forward rate constant associated with formation of dsDNA is much higher than the reverse rate constant then we find the expression for the zipping rate constant as that $k_Z = 1/\tau_Z \simeq k_p/L$ where we have defined $k_p = k_+ l_p$ (bases/s). On the other hand when $k_+ = k_-$ then the zipping will be a pure 1D diffusion process with the phenomenological diffusion coefficient as $D_\pm \simeq l_p^2 k_+$ (base$^2$/s) and subsequently $\tau_Z \simeq L^2/2D_\pm$. When the reacting c-ssDNA is repetitive with a sequence complexity of $c$ bases (here we have $c \in (1, L)$), then the observed zipping rate will be proportional to the number of repeats in that template c-ssDNA ($\rho = L/c$) and one obtains $\lim_{K_R \to 0} \tau_Z \simeq c/k_p$ i.e. the total zipping time will be directly proportional to the complexity of the reacting c-ssDNA molecules which is in line with the experimental observations (**6-7**). Using these results one can write down the expression for the total time that is required for the overall nucleation and zipping processes ($\tau_{NZ}$) as follows.

$$\tau_{NZ} = \tau_N + \tau_Z; \; \lim_{K_R \to 0} \tau_Z \simeq c/k_p; \; \lim_{K_R \to 1} \tau_Z \simeq Lc/2D_\pm; \; \tau_N = \lambda n^2/12 D_o; \; \lambda = L/n \tag{14}$$

## Calculation of overall renaturation rate

Using these values one can define the overall bimolecular collision rate associated with the complete formation of dsDNA from c-ssDNA in **Scheme I** as follows.

$$\bar{k}_S = \int_0^L k_S(n) p(n) dn; \quad k_S(n) = k_{fQ}/(1 + k_r \tau_{NZ}) \simeq k_{fQ}/(1 + nL/Y_A^2 + c/Y_B) \tag{15}$$

Here we have defined two important characteristic lengths $Y_A = \sqrt{12D_o/k_r}$ and $Y_B = k_P/k_r$. The length $Y_A$ describes the distance that is travelled by the initial incorrect contact before ci-ssDNA dissociates whereas $Y_B$ describes the distance travelled in the zipping reaction before ci-ssDNA dissociates into corresponding c-ssDNA molecules. Generally one observes that $Y_A > Y_B$. The probability density function connected with the 1D slithering lengths $n$ or its weighting function $p(n)$ can be calculated as follows. When the residence times ($\tau$) associated with the dissociation of single ci-ssDNA molecule is distributed as an exponential then one finds that $p(\tau) \propto e^{-k_r \tau}$ and subsequently one obtains $p(n) \propto ne^{-(n/Y_A)^2}$ which mainly originates from the fact that within the residence time $\tau$, the distance travelled by the incorrect contact through 1D diffusion dynamics can be anywhere in the interval $n \in (1, L)$ so that we obtain the transformation rule as $\tau = n^2/12D_o$. With this definition of the residence time of c-ssDNA in a ci-ssDNA configuration one can write down the expression for the distribution of slithering lengths $n$ as follows (**Fig. 2A**).

$$p(n) = 2ne^{-(n/Y_A)^2}/Y_A^2\left(1 - e^{-(L/Y_A)^2}\right) \simeq 2ne^{-(n/Y_A)^2}/Y_A^2 \tag{16}$$

Using the expression of $p(n)$ in **Eq. 15** and expanding $k_S(n)$ in a Maclaurin series one can obtain the following expression for the overall renaturation rate constant.

$$\bar{k}_S \simeq \left(k_{fQ}/(1 + c/Y_B)\right) \sum_{m=0}^{\infty} \Gamma(m/2 + 1)\left(-L/L_A (1 + c/Y_B)\right)^m \tag{17}$$

While deriving this equation without losing generality we have extended the limits of $n$ in the integration towards infinity since both $p(n)$ and $k_S(n)$ approach zero at this limit. Under certain conditions one can obtain the following approximation for the series in **Eq. 17**.

$$\bar{k}_S \simeq \left(k_{fQ} Y_B/c\right) \sum_{m=0}^{\infty} \Gamma(m/2 + 1)\left(-\rho Y_B/L_A\right)^m; \quad Y_A \gg Y_B; \quad (c/Y_B) \gg 1 \tag{18}$$

When the sequence complexity of the template c-ssDNA molecule is high enough and $(\rho Y_B/L_A) \ll 1$ then one can write down the leading zeroth order approximation ($m = 0$ in **Eqs. 17** and **18**) of the overall second order rate constant associated with the renaturation of relaxed c-ssDNA chains with equal lengths $L = l$ as follows.

$$\bar{k}_S \simeq k_{fG} Y_B/c = k_{fG} k_+/k_r c = k_t \chi_G k_p L/4 r_L k_r c \propto \sqrt{L}/c\eta; \quad \because r_L \propto 1/\sqrt{L}; \quad k_t \propto 1/\eta \tag{19}$$

Here $k_{sm} \simeq k_t \chi_G$ is the Smolochowski type three dimensional diffusion controlled collision rate limit corresponding to a situation where both colliding molecules are charged and same in size. When the conditions given in **Eq. 18** are true then **Eq. 19** suggests that the overall

Theory on the mechanism of DNA renaturation bimolecular collision rate constant associated with the renaturation reaction is directly proportional to the square-root of the length of c-ssDNA and inversely proportional to both sequence complexity of the reacting c-ssDNA molecules and viscosity of the reaction medium in line with the experimental observations. Experiments suggest that the scaling of renaturation rate constant with the length of c-ssDNA molecules that is given in **Eq. 19** is valid (**6**) only in the range of $L \sim 10^2\text{-}10^4$ bases. In this context our model suggest that the square root scaling of the renaturation rate on length will be valid only when the inequalities given in **Eq. 18** are true apart from the condition that $(\rho Y_B / L_A) \ll 1$ which may break down beyond certain values of the copy number ($\rho$) in the repetitive c-ssDNA.

**Role of cooperativity in renaturation kinetics**

However the scaling results given by **Eq. 19** works only for highly repetitive and relaxed conformational state of c-ssDNA ($L > c$) and it will break down at $L = c$ since at this point the scaling becomes as $\bar{k}_S \propto 1/\sqrt{L}$ which is not true. The main reason for this observation is that while calculating the zipping rate constant we have not considered the underlying cooperative effects. When c-ssDNA is highly repetitive then the zipping process can take place in parallel for all the $\rho = (L/c)$ number of short repetitive motifs. Under such conditions the cooperative effects will not be noticeable since the enhancement of renaturation process by the parallel-zipping will dominate over the underlying cooperative effects. This means that **Eq. 19** will be true only for repetitive c-ssDNA. When the reacting c-ssDNA molecules are non-repetitive and long enough then the probability of formation of an additional correct-contact in cc-ssDNA molecule that is undergoing zipping reaction will be directly proportional to the existing number of correct-contacts ($u$) and the probability of breakdown of an existing correct-contact will be directly proportional to the number of overhanging single stranded stretches of cc-ssDNA ($\beta$-$u$). This is true since the existing correct-contacts always stabilize newly formed correct-contacts and overhang single stranded regions of cc-ssDNA always try to destabilize the newly formed correct-contacts. Here one should note that we are dealing with the cooperative effects at a mesoscopic level within an independent and single renaturing cc-ssDNA molecule rather than at macroscopic level where the descriptive parameter of the renaturation process will be the number of mol-bases in ssDNA ($n_{SS}$) or dsDNA ($n_{DS}$) form rather than the number of correct-contacts in cc-ssDNA ($u$) as in the current context. At macroscopic level the rate of change in the number of mol-bases in ssDNA form in the process of zipping will be directly proportional to the number of mol-bases in ssDNA form as well as number of mol-bases in dsDNA form which results in a cooperative sigmoidal type time evolution of the renaturation process (**16-17, 28**) where the macroscopic kinetic rate equation will be written as $dn_{DS}/dt \propto n_{cc}(n_0 - n_{DS})n_{DS}$. Here $n_0$ is the initial concentration of mol-bases of ssDNA molecules in the system and $n_{cc}$ is the total number of cc-ssDNA molecules in the system. With this background the birth-death master **Eq. 10** can be rewritten to include the cooperative effects for renaturation of a single cc-ssDNA molecule as follows.

$$\partial_t P(u,t) = k_+(u-1)P(u-1,t) + k_-(\beta-u-1)P(u+1,t) - (k_+ u + k_-(\beta-u))P(u,t) \quad (20)$$

The mean first passage time $\tau(u)$ associated with evolution of the system from correct-contact $u = 1$ to complete dsDNA form with correct-contacts $u = \beta$ can be written as follows where $U(u)$ and other boundary conditions are defined as in **Eqs. 10-12**.

$$k_+ u U(u) - k_-(\beta-u) U(u-1) = -1;\ k_+ u \phi(u)[S(u) - S(u-1)] = -1;\ S(u) = U(u)/\phi(u) \quad (21)$$

Here the function $\phi(u)$ is defined as follows.

$$\phi(u) = \prod_{z=1}^{u} K_R(\beta-z)/z = (-K_R)^u \Gamma(u+1-\beta)/\Gamma(u+1)\Gamma(1-\beta)$$

Upon solving the difference equation **Eq. 21** for appropriate boundary conditions one obtains the following expression for the overall zipping time that is required for the formation of $u = \beta$ number of correct-contacts starting from $u = 1$ in the presence of cooperative effects.

$$\tau_Z = \sum_{u=1}^{\beta} K_R^u \left( \xi\, {}_2F_1([1,1],[2-\beta],-K_R^{-1}) + \phi\, {}_2F_1([1,u+1],[u+2-\beta],-K_R^{-1}) \right) \quad (22)$$

Here $_2F_1$ is the hypergeometric function and we have defined various parameters as follows.

$$\xi = (-1)^u \Gamma(u+1-\beta)/k_-\Gamma(u+1)\Gamma(1-\beta)(\beta-1);\ \phi = \Gamma(u+1-\beta)/K_R^{u+1} k_+ \Gamma(u+2-\beta)$$

The hypergeometric function of type $_2F_1$ is defined as follows.

$$_2F_1([a,b],c,z) = \sum_{m=0}^{\infty} z^m (a)_m (b)_m / m!(c)_m;\ (H)_P = \Gamma(H+P)/\Gamma(H)$$

To simplify the complicated expression for $\tau_Z$ in **Eq. 22** particularly for sufficiently large values of $\beta$ one can approximate **Eq. 20** by the following continuous type Fokker-Planck equation (FPE) (**36-38**).

$$\partial_t P(u,t) = -\partial_u (A(u) P(u,t)) + \partial_u^2 (B(u) P(u,t))/2 \quad (23)$$

Here the drift and diffusion coefficients can be written as follows.
$$A(u) = k_+ u - k_-(\beta-u);\ B(u) = k_+ u + k_-(\beta-u)$$

**Eq. 23** suggests that in the presence of cooperative effects the diffusion coefficient associated with the zipping dynamics ($D_\pm$) will be dependent on the number of correct-contacts. Using the backward FPE corresponding to **Eq. 23** one can obtain the mean first passage time associated with the evolution of the system from $u = 1$ to $u = \beta$ as follows.

$$\tau_Z \simeq (2/k_+) \int_1^{\beta} (\mathrm{H}(y)/\Phi(y)) dy;\ \mathrm{H}(y) = \int_1^y (\Phi(z)/(z+K_R(\beta-u))) dz \quad (24)$$

In this equation various functions and parameters are defined as follows.

$$\Phi(q) = e^{2\int_1^q p(z) dz};\ p(z) = A(z)/B(z) = (z - K_R(\beta-z))/(z + K_R(\beta-z))$$

Computational analysis of **Eqs. 22** and **24** suggests that in the limit as $K_R$ tends towards zero, the overall zipping time $\tau_Z$ approximately scales with $\beta$ as $1-e^{-2\beta}$. Upon defining the limit as $\lim_{K_R \to 0} \tau_Z = \tilde{\tau}_Z$ one can derive the following expression for the overall zipping time.

$$\tilde{\tau}_Z = (2/k_+) \int_1^{\beta} e^{-2y} (\mathrm{Ei}(1,-2) - \mathrm{Ei}(1,-2y)) dy;\ \mathrm{Ei}(a,z) = \int_1^{\infty} e^{-mz} m^{-a} dm \quad (25)$$

It seems from **Eq. 25** along with other computational analysis that the dependency of overall zipping time decreases with increasing $\beta$ in the presence of cooperative effects which can be demonstrated by the following limiting conditions.

$$\partial \tilde{\tau}_Z/\partial \beta \simeq 2e^{-2\beta}\left(\text{Ei}(1,-2) - \text{Ei}(1,-2\beta)\right)/k_+ ; \quad \lim_{\beta \to \infty} \partial \tilde{\tau}_Z/\partial \beta = 0 \quad (26)$$

From **Eqs. 24-26** one finds that when $K_R$ tends towards zero then in the presence of strong cooperative effects the zipping time of a non-repetitive c-ssDNA will be independent of the length of reacting c-ssDNA molecules especially for large values of $L$ as shown in **Fig. 2B**. Based on these observations we recover the observed scaling of overall bimolecular rate constant on the length of c-ssDNA molecules as $\bar{k}_S \propto \sqrt{L}/\eta$ for a non-repetitive c-ssDNA for which $L = c$ in **Eqs. 15** and **16** since $Y_B$ will be independent of sequence length and the number of copies will be $\rho = 1$.

## DISCUSSION

Understanding the mechanism of renaturation of c-ssDNA is one of the central topics in molecular biology and biological physics. Wetmur and Davidson (**6**) developed their model by assuming that the renaturation rate is directly proportional to the total phosphate concentration which is in turn directly proportional to the total number of mol-bases in ssDNA or dsDNA form. According to their model the overall second order rate constant associated with the renaturation of repetitive c-ssDNA can be written as $\bar{k}_S \propto k_N \rho$ where $\rho = L/c$ is the copy number of repetitive motifs in the entire c-ssDNA polymer. Here $L$ is the length of c-ssDNA and $c$ is the sequence complexity and the nucleation rate was assumed to scale with $L$ as $L^\alpha$ where $\alpha = -1/2$ due to the excluded volume effects associated with the interpenetration of c-ssDNA molecules that is essential for the nucleation reaction. As a result one obtains the scaling as $\bar{k}_S \propto \sqrt{L}/c$ where the proportionality constant was assumed to be the Smolochowski type bimolecular collision rate limit ($k_{sm}$) i.e. $\bar{k}_S = k_{sm}\sqrt{L}/c$. Here one can identify that $k_{sm} = (k_t \chi_G)$ of our model **Eq. 16** particularly for a relaxed conformational state of c-ssDNA that also includes the contributions from the electrostatic repulsions at the interface of colliding c-ssDNA molecules. Since in this model the nucleation is combined with incorrect-contact formation step one finds that $k_N = k_{sm}/\sqrt{L}$. The main arguments for this scaling result put forth by Wetmur and Davidson were viz. (a) the radius of gyration of c-ssDNA is directly proportional to its length and (b) the reaction radius associated with the collision between c-ssDNA molecules is independent of the radius of gyration of c-ssDNA chains since both these strands can interpenetrate freely upon their collision. Though the assumption (a) is a right one for Gaussian chain polymers there are several questions with the assumption (b) since the reaction radius always depends on the sum of the radii of gyration of reactant molecules. In this context our detailed model clarifies the origin of such scaling in the renaturation phenomenon. It is clear from our theory that the rate constant associated with the formation of initial incorrect-contact is directly proportional to the square-root of the length of the reacting c-ssDNA molecule.

One should note that it is very difficult to identify and isolate a nucleated cc-ssDNA molecules since they are indistinguishable from the zipping cc-ssDNA molecules. Therefore it is more appropriate to combine the nucleation step with the zipping step rather than with the incorrect-contact formation step as in case of Wetmur-Davidson model. Other issues in their model are such as the breakdown of scaling at $L = c$ arises because the underlying cooperative effects in the long and non-repetitive c-ssDNA are not considered in their model as in case of our **Eq. 16**. Further upon extrapolating towards the limit $L = 1$ base (so that the complexity becomes as $c = 1$ base and the radius of gyration of c-ssDNA molecules $r_L \approx 1$ base) Wetmur-Davidson model predicted that $\bar{k}_S = k_{sm}$. However experimental observations suggested that the extrapolated bimolecular collision rate constant associated with the renaturation reaction was ~$10^3$ times lower than the Smolochowski type diffusion controlled bimolecular collision rate limit. On this basis they in turn discarded the possibility of diffusion control in the kinetics of renaturation of c-ssDNA molecules. In this context **Eq. 16** of our model suggests an approximate expression for the extrapolation intercept as $k_{sm} \simeq k_t \chi_G \varepsilon$ from which one finds that $\varepsilon = \left(k_p/4k_r\right) \simeq 10^{-3}$. One should note that **scheme I** of **Fig. 1** is still valid with zero nucleation and zipping times and the steric factor $\varepsilon$ mainly accounts for the geometric constraints associated with the bond formation between nitrogen bases A-T or G-C of the colliding single nucleotides in the limit $L = 1$ and $c = 1$. Here one should note that in our model the square-root dependency of renaturation rate on the length of c-ssDNA molecules mainly originates from the fact that the radius of gyration of c-ssDNA molecules scales with length as $r_L \propto L^\alpha$ where $\alpha = ½$ which is valid only for a Gaussian type polymers in a theta solvent. It seems that the error introduced by this assumption in the exponent is within the experimental error range (**6**).

Although the dissociation rate (here it is $k_r$) constant increases exponentially with temperature (**39-42**) there are several controversies exist on the dependency of renaturation rate constant on increasing temperature. Some experimental studies established (**39-40**) a decrease in the renaturation rate constant with increase in temperature and some other studies have shown an increase in the renaturation rate constant with increase in temperature (**8, 41**). In general it seems that the temperature dependency of the renaturation rate constant is of non-Arrhenius one and non-monotonic type (**42**). Simulation studies suggested that there exists an optimum temperature at which the renaturation rate constant is a maximum (**29**). In our model the incorrect-contact formation, nucleation and zipping steps are all influenced by the rise in temperature in a complicated way. Actually in **Eq. 16**, the Smolochowski bimolecular collision rate constant depends on temperature as $k_t = \left(8k_BT/3\eta\right)$ where we assume that viscosity of the medium is not changing much in the range of temperature variation and the dissociation rate scales with temperature as $k_r = k_r^0 e^{-\omega/k_BT}$ in line with transition state theory where $\omega$ is the free energy barrier associated with the dissociation of ci-ssDNA complex. The rate constant associated with the microscopic zipping ($k_P \sim l_p k_+$) is connected with the microscopic diffusion coefficient $D_\pm \sim l_p^2 k_+$ associated with the zipping reaction. Apart from these the dimensionless parameter $\chi_Q$ corresponding to the overall electrostatic repulsions and the shielding effects of solvent ions at the interface of ci-ssDNA molecules also depends on the temperature as given in **Eqs. 5-6**. It seems that the non-Arrhenius type kinetic behaviour arises as a consequence of a complicated interplay between increase in the rate of incorrect-contact formation and combined effects of increase in the dissociation rate constant and microscopic zipping rate constant as the temperature increases from low to high values. Upon noticing that $\bar{k}_S \propto k_B T e^{\omega/k_BT}$ one finds that $\ln \bar{k}_S$ will be a maximum approximately at $T \sim \omega/k_B$ as observed in the simulation studies (**29**).

Sikorav et.al (**27**) suggested a Kramer's type expression for the bimolecular nucleation rate constant where the scaling dependency of the overall bimolecular collision rate associated with renaturation on the length of c-ssDNA mainly originates from the entropic component of free energy barrier. However in this model the reaction coordinate and origin of free energy barrier associated with the nucleation and zipping are not clearly defined. Further the exact mechanism of formation of nucleation sites is not clearly explained. On the other hand as correctly pointed out by them, one cannot explain all the experimental observations related to the entire process of renaturation of c-ssDNA molecules with purely diffusion-controlled formalism or transition state theoretical framework. From our model we can conclude that the incorrect contact formation step is a pure three dimensional diffusion controlled collision rate processes whereas both nucleation and zipping steps involve a sequence of several microscopic crossings of free-energy barriers as well as one dimensional diffusion type slithering dynamics on a linear lattice.

Condensed conformational state of c-ssDNA polymers is one more cause for the breakdown of the scaling of renaturation rate on the length of c-ssDNA that is given in **Eq. 16**. When the colliding c-ssDNA molecules are in condensed conformational state then the rate constant associated with the incorrect-contact formation step will be independent of the length of c-ssDNA when $L = l$. Under such conditions the overall second order rate constant associated with the renaturation of repetitive c-ssDNA chains will be inversely proportional to the sequence complexity. Here one should note that while deducing these facts we have not considered the condensation of both c-ssDNA molecules together which is known to enhance the overall renaturation rate over several orders of magnitude as in case of renaturation in phenol-water interface (**43**). Under such co-condensation of both strands of c-ssDNA the rate of incorrect-contact formation is very large and the dissociation rate will be very small and the rate limiting steps are the nucleation and zipping ones.

## CONCLUSIONS

Renaturation (or hybridization) of complementary single strands of DNA is an important phenomenon in molecular biology and biological physics. Understanding the kinetic mechanism of renaturation is very much useful to further understand the winding-unwinding dynamics of double stranded DNA under both *in vitro* and *in vivo* conditions. Here we have developed a stochastic dynamics based model on the DNA renaturation phenomenon to explain various scaling behaviours of renaturation rate. According to our model there are at least three steps in the renaturation process viz. incorrect-contact formation, stochastic nucleation and zipping. Most of the earlier two-state models combined nucleation with incorrect-contact formation step. We argue that it is considerably meaningful when we combine the nucleation with the zipping since nucleation is the initial step of zipping. Incorrect-contact formation step is a pure three-dimensional diffusion controlled collision process whereas nucleation involves several rounds of one-dimensional slithering dynamics of one single strand of DNA on the other complementary strand in the process of searching for the correct-contact and initiate nucleation. Upon nucleation, the stochastic zipping follows to generate a fully renatured double stranded DNA.

It seems that the square-root dependency of the overall renaturation rate constant on the length of reacting single strands originates mainly from the geometric constraints in the diffusion controlled incorrect-contact formation step. On the other hand the inverse scaling of the renaturation rate with the sequence complexity originates from the stochastic zipping which involves several rounds of crossing of free-energy barrier at microscopic level. When the sequence of renaturing single strands of DNA is repetitive with less complexity then the cooperative effects will not be noticeable since the parallel zipping will be a dominating enhancement factor. However for DNA strand with high sequence complexity and length one needs to consider the cooperative effects both at microscopic and macroscopic levels to explain various scaling and kinetic behaviours of the overall renaturation rate.

FIGURES CAPTIONS

FIGURE 1

**A**. Three basic steps in the renaturation of complementary single strands of DNA (c-ssDNA) are viz. incorrect-contact formation, nucleation and zipping. Incorrect-contact formation (ci-ssDNA) is purely a three dimensional (3D) diffusion controlled collision rate process (I) where the rate constant associated with the formation of incorrect-contact scales with the length of colliding c-ssDNA molecules in a square root manner and it scales with the solvent viscosity in an inverse manner. Nucleation involves a one dimensional (1D) slithering dynamics (II) of one strand on the other strand of ci-ssDNA in the process of searching for correct-contact (cc-ssDNA). Upon finding the correct-contact and forming the nucleus zipping of cc-ssDNA step (III) follows. Since nucleated cc-ssDNA is indistinguishable from zipping one, it is more appropriate to combine the nucleation with the zipping rather than with the incorrect-contact formation step as shown in **Scheme I** where both the nucleation and zipping are coupled stochastic processes. Conformational state of the reacting c-ssDNA molecules seems to significantly affect the reaction mechanism and scaling relationships associated with the overall renaturation rate.

**B**, **C**. We can model the c-ssDNA chains as clusters of nitrogen bases so that the overall bimolecular rate associated with the formation of incorrect contacts between spatially distributed base-clusters of c-ssDNAs is proportional of the product of concentrations of the total nitrogen bases in c-ssDNA molecules. The cylindrical surface area $S_Q \sim 2\pi r_D Q$ of a c-ssDNA molecule will be confined within the spherical solvent shell with surface area $SS_Q \simeq 4\pi r_Q^2$ ($Q = L$ for template and $Q = l$ for probe) where $r_Q$ is the radius of gyration of the respective c-ssDNA molecule. Under strongly condensed state of c-ssDNA one finds that $SS_Q < S_Q$ (**C**) and when the DNA polymer is in a relaxed state (**B**) then one find that $SS_Q > S_Q$. At a coarse grained level one can model the bases of c-ssDNA as a chain of spherical beads with radius $r_D$. Under relaxed conformational state all these nitrogen base beads are distributed on the surface of the spherical solvent shell that covers a c-ssDNA molecule (**B**). Under condensed conformational state of c-ssDNA molecules significant fraction of nitrogen base beads will be inaccessible to the inflowing c-ssDNA molecules since they are buried inside the matrix of condensed c-ssDNA (**C**).

FIGURE 2

**A**. Probability density function associated with the one dimensional slithering length (*n* measured in base) of ci-ssDNA in the process of searching for the correct-contact as given in **Eq. 16** for different values of the characteristic length $Y_A = \sqrt{12 D_o / k_r}$ from 10 to 100 bases where $D_o$ (base$^2$s$^{-1}$) is the one dimensional diffusion coefficient associated with the slithering dynamics and $k_r$ is the dissociation rate constant connected with ci-ssDNA.

**B**. Zipping time ($\tau_Z$, measured in seconds) in the presence of cooperative effects. Here sequence complexity (*c*) is same as that of the length (*L*) of c-ssDNA i.e. $c = L$. The number of correct-contacts $\beta = L/l_p$ is a dimensionless quantity where $l_p = 1$ base and $L$ is the length of the reacting c-ssDNA. Green solid line is calculation from **Eq. 22** and blue solid line is calculation from **Eq. 24**. Here we have set $K_R \sim 10^{-6}$ and $k_+ \sim 1$ s$^{-1}$. Red solid line is the derivative of zipping time with respect to $\beta$ as in **Eq. 26** which shows that the value of the derivative of overall zipping time with respect to $\beta$ is $< 10^{-2}$ when $\beta > 10^2$. These plots suggest that when $K_R$ tends towards zero, the overall zipping time of a non-repetitive and long c-ssDNA will be independent of the length of the reacting c-ssDNA molecules. Zipping time of a repetitive c-ssDNA with a sequence complexity of *c* bases increases linearly with *c*.

FIGURE 1

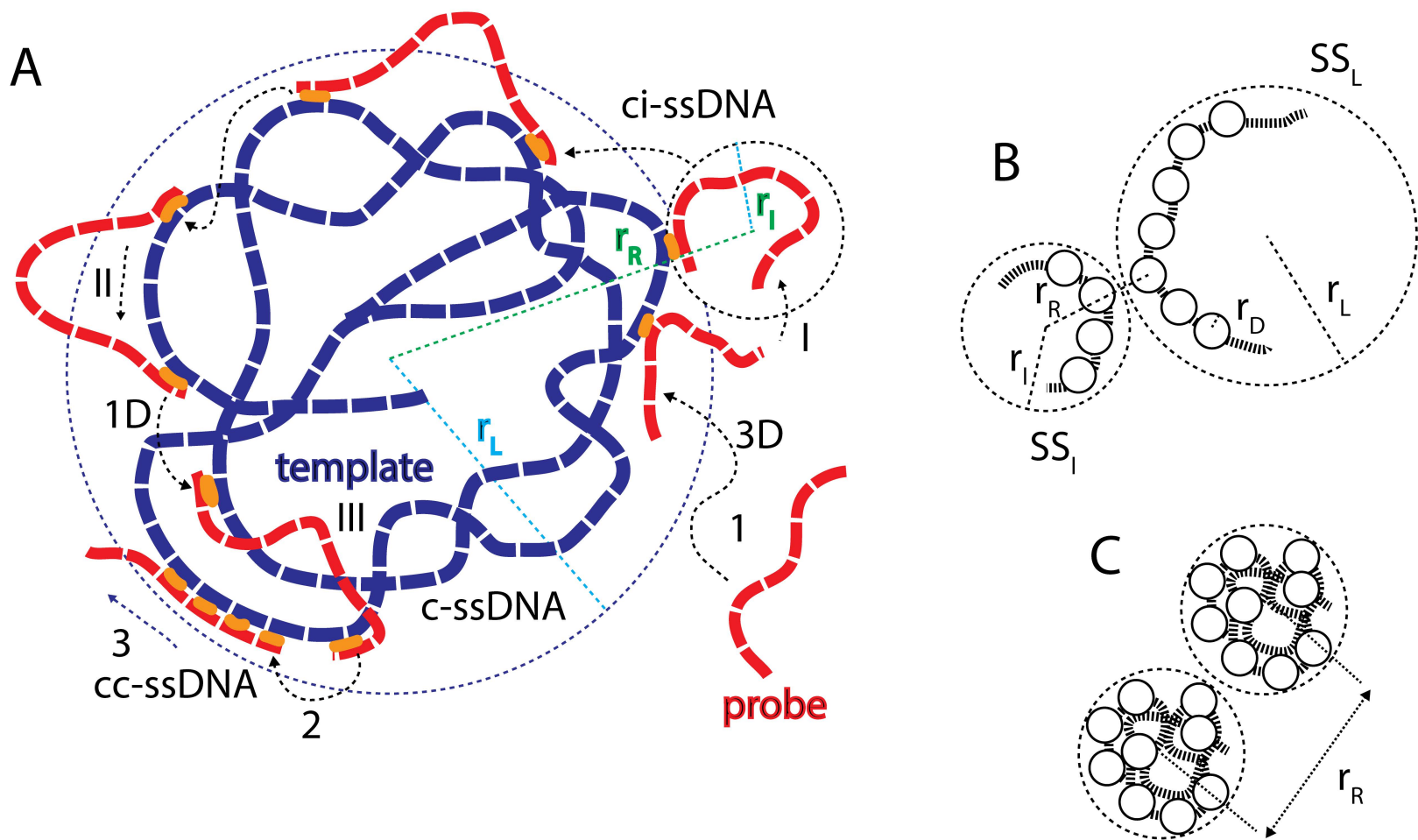

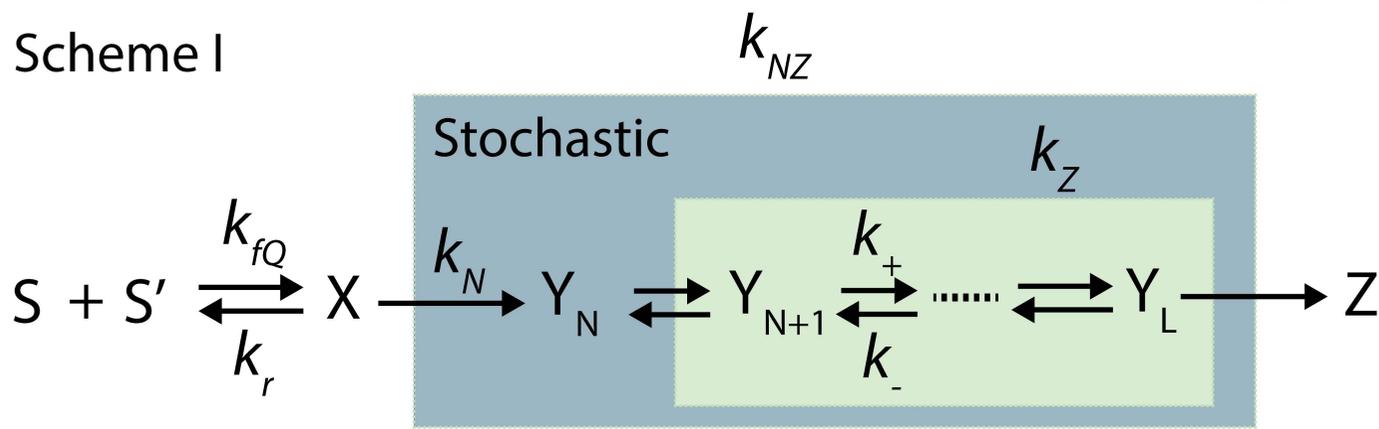

# FIGURE 2

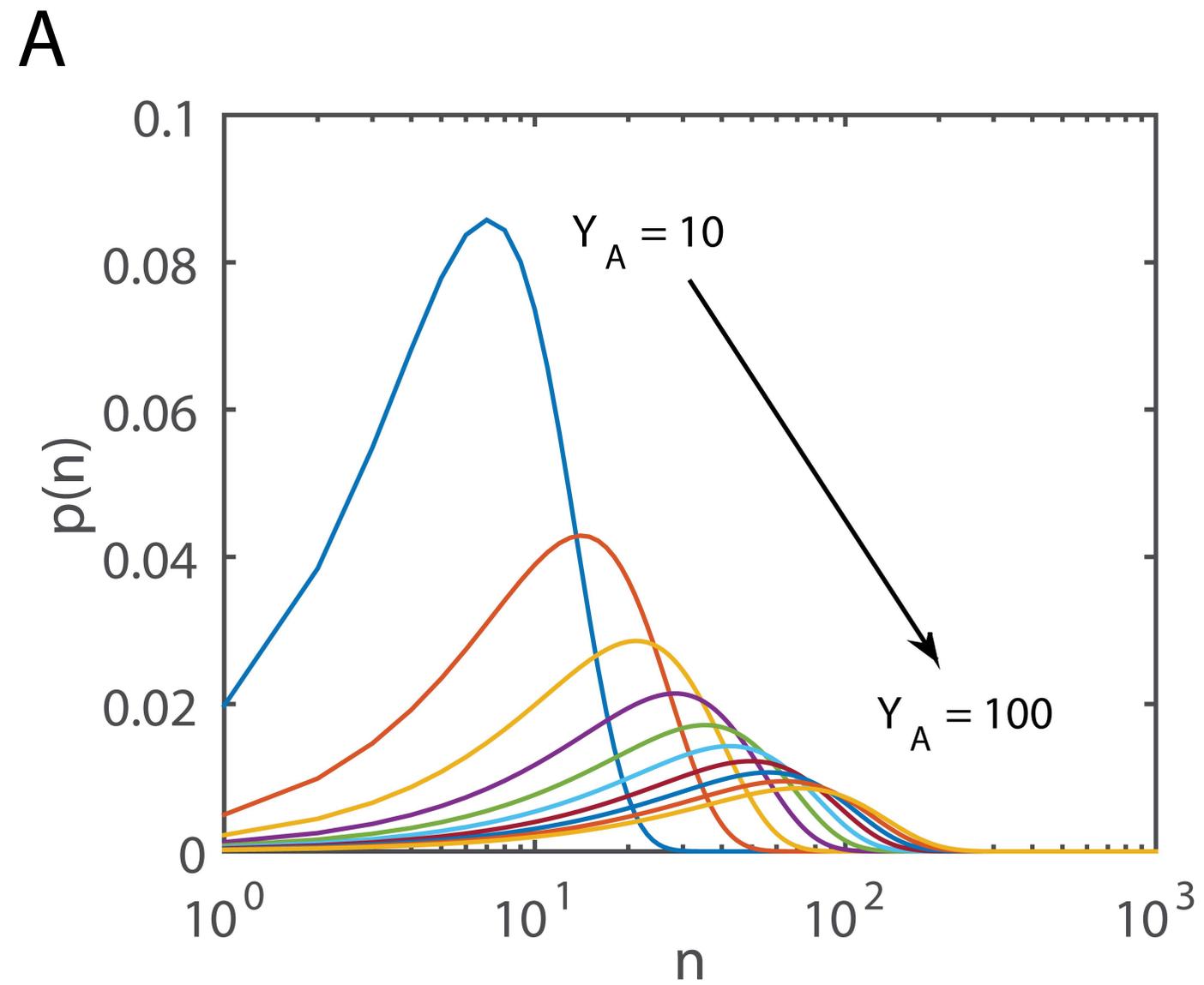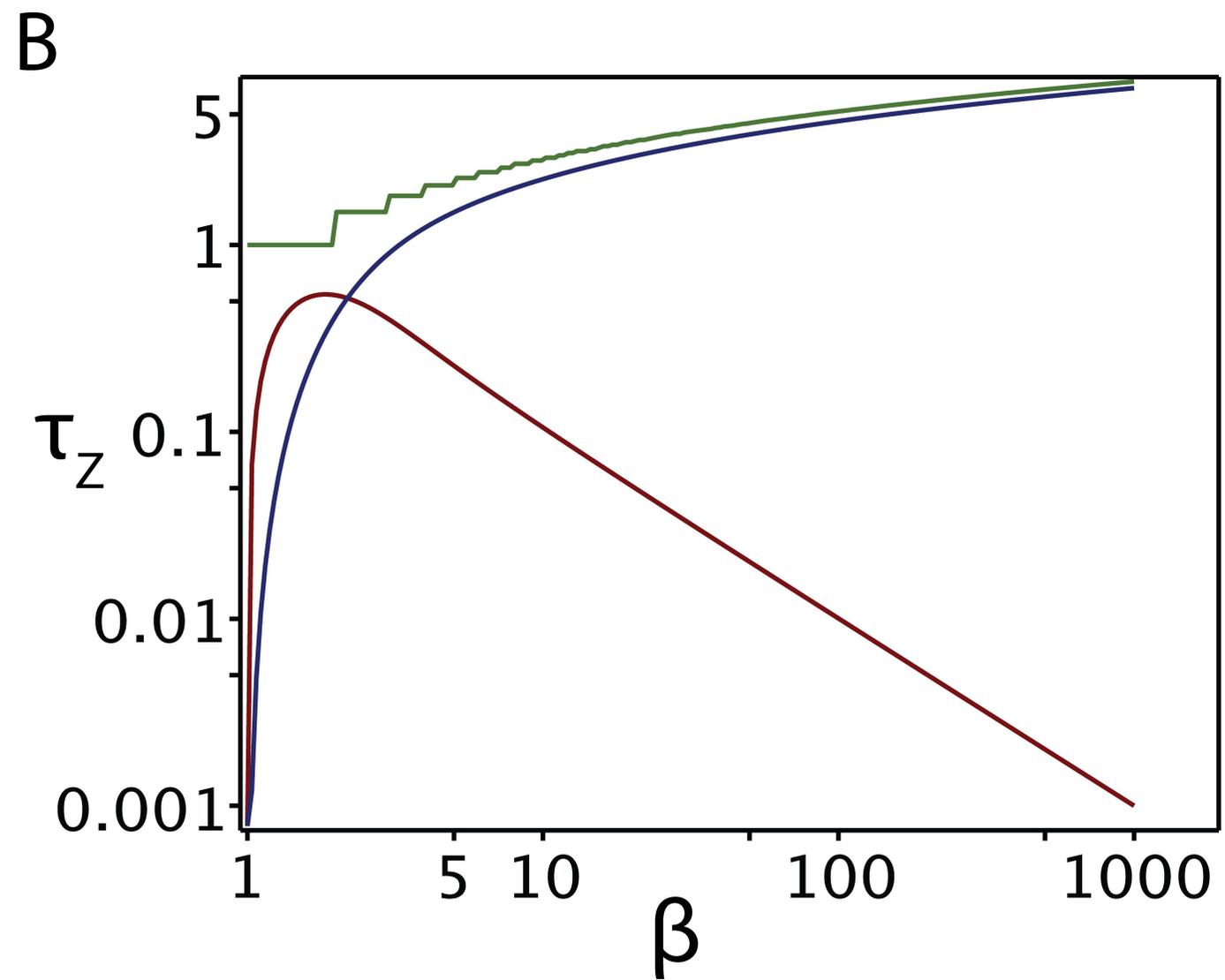